\documentstyle[psfig]{l-aa}

\begin{document}

 \title{Is $\gamma$-ray absorption by induced electric fields important 
        in the pulsar magnetospheres?}
 \author{Zheng Zheng\inst{1,2,3},
         Bing Zhang\inst{1,3}, G. J. Qiao\inst{4,1,3}}
 \offprints{Z. Zheng.}
 \institute{Department of Geophysics, Peking University, Beijing, 100871,
            P. R. China
  \and 	   Beijing Astronomical Observatory, Chinese Academy of Sciences, 
           Beijing, 100012, P. R. China
  \and     CAS-PKU Joint Beijing Astrophysics Center, Beijing, 100871,
           P. R. China
  \and     CCAST (World Laboratory) P.O.Box 8730, Beijing, 100080, P. R. 
           China\\
           E-mail: ZZ zhengz@bac.pku.edu.cn; BZ zb@bac.pku.edu.cn;
           GJQ gjn@pku.edu.cn }
 \thesaurus{00.00.0, 00.00.0}
 \date{Received date ; accepted date}
 \maketitle
 \markboth{Z.Zheng, B.Zhang, \& G.J.Qiao: Is $\gamma$-ray absorption by 
induced electric fields important?}{}


\begin{abstract}  

Although the unified formula for $\gamma$-ray absorption process involving 
both the magnetic field and a perpendicular electric field derived by 
Daugherty \& Lerche (1975) is correct, we argued in this paper that their 
conclusion that the induced electric fields are important in the pair 
formation process in the pulsar magnetospheres is wrong and misleading. The 
key point is that usually the direction of a $\gamma$ photon at the emission 
point observed in the laboratory frame should be $(v/c, 0, [1-(v/c)^2]^{1/2})$ 
rather than $(0, 0, 1)$, where $v$ is the co-rotating velocity. This 
emission direction is just the one which results in zero attenuation 
coefficient of the $\gamma$ photon. Calculation shows that after the photon 
has moved a distance, its direction lead to the result that the induced 
electric field is also of minor importance. Thus only $\gamma-B$ process is 
the important mechanism for the pair production in the pulsar 
magnetospheres. The implications of the modification by ejecting the induced 
electric field are also discussed.

\keywords{Pulsars: general -- Pair production}

\end{abstract}

\section{Introduction}

Pair production process plays an important role in pulsar physics. It 
is not only a necessary process for the multiplication of the particles
to account for emissions of different bands from pulsars (e.g. Sturrock 
1971; Ruderman \& Sutherland 1975; Arons \& Scharleman 1979; Arons 1983; 
Cheng, Ho, \& Ruderman 1986), but also an important mechanism to absorb 
$\gamma$-rays produced in the pulsar magnetospheres, especially near the 
polar cap region (e.g. Hardee 1977; Harding, Tademaru, \& Esposito 1978; 
Harding 1981; Daugherty \& Harding 1982, 1996; Zhao {\it et al.} 1989; 
Lu \& Shi 1990; Lu, Wei, \& Song 1994; Dermer \& Sturner 1994; Sturner, 
Dermer, \& Michel 1995; Wei, Song, \& Lu 1997). Furthermore, the way by 
which the $\gamma$-rays are absorbed is also the key factor to limit 
the parameters of the inner magnetospheric accelerators of pulsars (e.g. 
Ruderman \& Sutherland 1975, hereafter RS75; Zhang \& Qiao 1996; Qiao \& 
Zhang 1996; Zhang {\it et al.} 1997a; Zhang, Qiao, \& Han 1997b, hereafter 
ZQH97b). 

Pair formation in intense magnetic fields ($\gamma-B$ process) has been 
studied explicitly by different authors (e.g. Erber 1966; Tsai \& Erber 
1974; Daugherty \& Harding 1983, Rifert, M\'{e}sz\'{a}ros \& Bagoly 1989), 
and its importance in pulsar physics was first pointed out by Sturrock 
(1971). Daugherty \& Lerche (1975, hereafter DL75) first dealt with the 
case involving a relatively weaker electric field perpendicular to the 
magnetic field (${\bf E}^2-{\bf B}^2\le0$, ${\bf E\cdot B}=0$), and came to 
a unified formula of the attenuation coefficient of the $\gamma$ photons. 
The more general case involving both the perpendicular and the parallel 
components of the electric field with respect to the magnetic field 
($E_\perp$ and $E_\parallel$) was presented by Daugherty \& Lerche (1976) 
and Urrutia (1978). In the specific case of pulsars, although $E_\parallel$ 
is usually sufficiently small so that its effect is negligible, $E_\perp$ 
induced by the fast spin of the neutron stars was demonstrated to be very 
important in pair formation process by DL75. This leads many authors to 
take this effect seriously into account in their studies (e.g. Hardee 1977; 
 Lu \& Shi 1990; Lu, Wei, \& Song 1994; Qiao \& Zhang 1996). 

In this paper, we'll argue that although the unified formula of DL75 is 
correct, their conclusion that the induced electric fields are important in 
the pair formation process in the pulsar magnetospheres is wrong and 
misleading. The detailed argument is presented in Section 2 and Section 3. 
Finally, we discuss the possible implications of this modification.

\section{The role of the induced electric fields in pair formation
at the emission point}

The attenuation coefficient of converting a $\gamma$ photon into 
electron-positron pairs in a pure strong magnetic field with a perpendicular 
component $B_\perp$ is expressed as
$$\zeta=0.23c{\alpha_0\over\lambda_e}{2mc^2\over E_\gamma}\chi
\exp(-{4\over 3\chi}), \eqno(1)$$
with $\chi={E_\gamma\over 2mc^2}{B_\perp \over B_c}$ under the condition of 
$\chi\ll 1$ (Erber 1966). Here $\alpha_0=e^2/\hbar c$ denotes the fine 
structure constant, $\lambda_e=\hbar/mc$ is the reduced Compton wavelength 
of the electron, $B_c=m^2c^3/e\hbar=4.414\times10^{13}$G is the critical 
magnetic field, and $E_\gamma$ is the energy of the $\gamma$ photon.

If a relatively weaker electric field perpendicular to the magnetic field 
(i.e. ${\bf E}^2-{\bf B}^2\le0$, ${\bf E\cdot B}=0$) exists, the attenuation 
coefficient can be derived by performing a Lorentz transformation (DL75). 
With the positive $y$ and $z$-axes being the direction of ${\bf E}$ and 
${\bf B}$, respectively, this attenuation coefficient, which is a function of 
the direction cosines ($\eta_x$, $\eta_y$, $\eta_z$) of the photon, is 
expressed by DL75 (see their Eq.(9)). Note that the positive $x$-axis is just 
the moving direction of a frame with a ``drifting'' velocity of
${\bf v}=c{{\bf E\times B}\over B^2}$, in which the electric field vanishes 
completely. This velocity is also the ``co-rotating velocity'' of the 
particles in a pulsar's magnetosphere when the magnetic axis is aligned with 
the rotational axis. 

DL75's derivation of the general formula (their Eq.(9)) is by all means 
correct. For $(\eta_x, \eta_y, \eta_z)=(E/B, 0, \pm[1-E^2/B^2]^{1/2})$
(their {\it case (d)}), they derived $\zeta\equiv 0$, which means that in 
two special directions, the $\gamma$ photons can traverse the 
electromagnetic field freely without being absorbed at all. Another 
special case is $(\eta_x, \eta_y, \eta_z)=(0, 0, 1)$ (their {\it case (e)}), 
i.e. the $\gamma$ photon is moving along the direction of the magnetic field. 
And they obtained
$$\zeta=0.23c{\alpha_0 \over \lambda_e}{2mc^2\over E_\gamma}\chi (1-E^2/B^2)
\exp (-{4\over 3\chi}), \eqno(2)$$
where $\chi={E_\gamma\over 2mc^2}{E\over B_c}$. It just like that $E$ in 
Eq.(2) has replaced $B_\perp$ in Eq.(1) apart from a multiplicative factor 
$(1-E^2/B^2)$. 

In a pulsar magnetosphere, when we observe in the laboratory frame, a 
spin-induced electric field satisfying ${\bf E}+{1\over c}{\bf v\times B}=0$ 
should exist to fulfill the force-free condition (Goldreich \& Julian 1969),
where ${\bf v}$ is the co-rotating velocity at a given point in the 
magnetosphere. This electric field is perpendicular to $B$ and is much 
smaller than $B$ since $v\ll c$ in the pulsar polar cap region. So DL75's
formula can be used to deal with this problem. DL75 argued that the photon's 
direction should be nearly parallel to $B$ in the pulsar magnetosphere, so 
they came to the conclusion that their {\it case (e)} is proper and the 
induced electric field plays a considerable role in the pair production 
process, especially for those rapidly spinning pulsars.

But this argument is unfortunately wrong, since it results in very unnatural
conclusions. First, consider a special magnetic field configuration 
(although it does not exist at all in nature) with straight field lines  
co-rotating with the magnetosphere. In the co-rotating frame where the 
electric field vanishes, a particle will move strictly along the field line 
and hence, emit a $\gamma$ photon by a certain mechanism in the direction 
(0, 0, 1). The $\gamma$ ray cannot be absorbed since there is no 
perpendicular magnetic field component at all. But if one observes the same 
process in the laboratory frame, according to DL75, Eq.(2) shows that the 
absorption is quite severe since there exists a very strong rotation-induced 
electric field. We see contradictory pictures in two different observer's 
frames. Secondly, in a curved magnetic field, a $\gamma$-ray can travel a 
distance until $B_\perp$ achieves a sufficient value to absorb it. But 
according to Eq.(2), a $\gamma$-ray can be directly absorbed severely at the 
very position it is produced. This means that in a rapidly rotating pulsar 
magnetosphere, the relatively high energy $\gamma$ photons can hardly be 
formed at all. This is also quite unnatural. What is wrong?

The problem lies in the DL75's misusing (0, 0, 1) as the photon direction in 
the laboratory frame. In a strong magnetic field, the very short lifetime on 
the higher Landau energy levels makes an electron almost always stay in its 
ground state, so it is described that an electron moves along the field 
lines and emits photons at the tangent moving direction (e.g. the curvature 
radiation or the inverse Compton scattering). But this picture is strictly 
correct ONLY in the frame where there is no perpendicular electric field, or 
in the ``co-rotating'' frame for a aligned pulsar. If the observer moves 
with respect to the magnetic field (e.g. in the laboratory frame) so that an 
induced electric field exists, the directions of the field line, of the 
electron, and of the photon are all different from each other since they 
obey different transformation laws.

Suppose two inertia frames: frame  $S^\prime$, the instantaneous co-rotating 
frame where the electric field vanishes so that the three directions are 
aligned; and frame $S$, the laboratory frame where the three directions may 
be misaligned (denoted by the angles between them and the positive x-axis, 
$\theta_B$, $\theta_u$ and $\theta_\gamma$). Frame $S$ moves at a relative 
speed $-{\bf v}$ with respect to frame $S^{\prime}$, where ${\bf v}$ is the 
co-rotating velocity. The positive $x$-axis is defined as the direction of 
${\bf v}$. The Lorentz transformation of the electromagnetic field then 
results in (in order to be simple, we assume the direction of 
${\bf B}^\prime$ is in $x^\prime-z^\prime$ plane) 
$$\tan\theta_B=\gamma_r\tan\theta_B^\prime, \eqno(3)$$
where 
$\gamma_r=1/\sqrt{1-v^2/c^2}$ 
and the subscript ``r'' tells the Lorentz factor of the rotation from that 
of the particle. And the velocity Lorentz transformation comes to
$$\tan\theta_u={1\over\gamma_r}{\sin\theta_u^\prime\over\cos\theta_u^\prime 
  +{v\over u^{\prime}}},\eqno(4)$$
where $u^{\prime}$ is the velocity of the particle along the magnetic field 
line in the frame $S^{\prime}$. The aberration effect makes
$$\tan\theta_\gamma={1\over\gamma_r}{\sin\theta_\gamma^\prime\over
  \cos\theta_\gamma^\prime +{v\over c}}.\eqno(5)$$
Although $\theta_B^\prime= \theta_u^\prime=\theta_\gamma^\prime$  in the 
frame $S^\prime$, usually $\theta_B\neq\theta_u\neq\theta_\gamma$ in the 
frame $S$. Physically, since an induced electric field ${\bf E}$ exists in 
the laboratory frame $S$, a ``drifting'' velocity component 
${\bf v}=c{{\bf E\times B} \over B^2}$ will be added to the electrons 
besides the component along the magnetic field ${\bf B}$.

\begin{figure}[t]
 \centerline{\psfig{figure=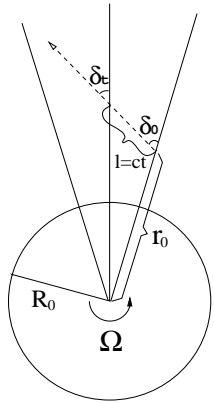,angle=270,height=5.353cm,width=4.5cm}}
 \caption[]{The geometry of the photon moving direction with respect to
            the magnetic field configuration in an aligned rotator, viewed
            from the magnetic pole, i.e. the rotational pole.}
 \label{Figphoton}
\end{figure}

In the case of an ``aligned rotator'' whose magnetic and rotational axes are 
in the same direction, the relative velocity of the two frames is 
perpendicular to the magnetic field line in the co-moving frame, i.e.  
$\theta_B^\prime=\theta_u^\prime= \theta_\gamma^\prime =\pi/2$. Using Eqs.
(3,5), we can obtain $\theta_B=\pi/2$ and the photon's direction in frame 
$S$ as $(\eta_x, \eta_y, \eta_z)=(\sin\delta, 0, \cos\delta)=(v/c, 0,
[1-(v/c)^2]^{1/2})$ rather than $(0, 0, 1)$, where $\delta$ is the angle 
between the direction of $\gamma$-ray and that of the field line in frame 
$S$ (i.e. $\theta_B-\theta_\gamma$). Note this direction is just the 
direction $(E/B, 0, [1-(E/B)^2]^{1/2})$ (for an aligned rotator ${\bf v}$, 
${\bf E}$ and ${\bf B}$ are perpendicular to each other). This is just the 
{\it case (d)} rather than the {\it case (e)} in DL75, so that $\zeta=0$ is 
satisfied, i.e. the induced electric field plays no role in the $\gamma$-ray 
absorption process at the emission point since the emission is produced at a 
preferred direction.

\begin{figure}[t]
 \centerline{\psfig{figure=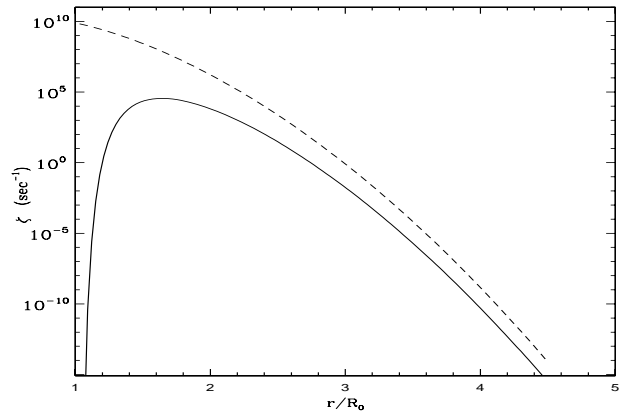,angle=0,height=5.4cm,width=8.5cm}}
 \caption[]{The attenuation coefficient curve (the solid line) of a $\gamma$ 
            photon as a function of the distance it travels (see detailed 
            description in the text). The curve of case $(0, 0, 1)$ is also 
            plotted by the dashed line. Parameters adopted: pulsar period 
            $P=0.1$s, surface magnetic field $B=10^{12}$G, height of emission 
            point $r_0=1.01R_0$, energy of the photon $E_\gamma\sim 10^4$Mev.}
 \label{Figcoef}
\end{figure}

The two ``unnatural conclusions'' get away now. In the ``straight field''
configuration, the $\gamma$ photon will not be absorbed in the laboratory 
frame, either, since it is just emitted in the absorption-free direction. 
So no contradiction lies in the two frames. Furthermore, no energy 
restriction on pair production is performed anywhere in the pulsar 
magnetosphere because a $\gamma$-ray will never be severely absorbed before 
it moves a distance for $B_\perp$ to achieve a sufficient value to absorb it. 

In the polar cap region of a pulsar, the co-rotating velocity is much smaller
than the speed of light, so that $\delta=\sin^{-1} (v/c)$ is very small and 
usually we can still regard the three directions described above to be the 
same. But in discussing the pair formation process, the attenuation 
coefficient $\zeta$ is very sensitive to the incident angle. It is just this 
very small deviation of the incident angle that really counts. Now another 
question arises: Propagation makes a $\gamma$ photon encounter other bunches 
of field lines and changes the incident angle. Will the induced electric 
field be important at this time?

\section{The role of the induced electric fields in pair formation
after the photon moves a distance}

According to the above discussions, a $\gamma$ photon will be initially 
emitted at an incident angle $\delta_0=\sin^{-1}(v/c)$ in the laboratory 
frame for an aligned rotator, where $v$ is the co-rotating velocity at the 
emission point. Neglecting the general relativity effect, the photon will 
move along a straight line and intersect other bunches of the magnetic field 
lines. The incident angle of this $\gamma$ photon gets smaller and smaller, 
and asymptotically merges to zero at infinity where the direction $(0, 0, 1)$ 
is usable, if we again assume the magnetic field lines are straight and are 
perpendicular to the rotating axis (i.e. there is no 
field-line-curvature-caused $B_\perp$ at all after the photon moves a 
distance, see Fig.1). The magnetic field is assumed to be also declined as 
$r^{-3}$. Our adopting this unrealistic but simple magnetic field 
configuration is just for the sake of examining the importance of the  
propagation-caused absorption effect.

Suppose the $\gamma$ photon with $E_\gamma\sim 10^4$Mev is emitted at a 
height of $(r_0-R_0)$ from the neutron star surface ($R_0$ is the radius 
of the neutron star). After it travels a distance $l=ct$ ($t$ is the travel 
time), the incident angle changes to $\delta_t$.
Submitting $(\eta_x,\eta_y, \eta_z)=(\sin\delta_t, 0, \cos\delta_t)$ into 
DL75's Eq.(9), we can get the attenuation coefficient of the photon as a 
function of the travel distance (see Fig.2). The extreme case of $(\eta_x, 
\eta_y, \eta_z)=(0, 0, 1)$ is also marked. From Fig.2 we can see that the 
induced electric field is also of minor importance, since the attenuation 
coefficients are always at least two order-of-magnitudes smaller than that 
of the case (0, 0, 1). 

 
\section{Conclusion and discussions}

Although the unified formula for the pair production process involving both
the magnetic field and the perpendicular electric field derived by DL75 is 
correct, we have argued in this paper that their conclusion that the induced 
electric fields are important in the pair formation process in the pulsar 
magnetospheres is wrong and misleading at least for the ``aligned rotator''
case. At the emission point, the photon emitted by a certain mechanism (e.g. 
the curvature radiation or the inverse Compton scattering) just moves along 
the very direction in which the attenuation coefficient is zero. Considering 
the propagation of the photon, we found that this rotation-induced electric 
field still plays a minor role in the $\gamma$-ray absorption process in the 
polar cap region of a pulsar. 

For the general case of an ``oblique rotator'' in which the magnetic and the 
rotational axes are misaligned, 
the co-rotating velocity is no more perpendicular to the magnetic field so 
that $\theta_B^{\prime}$ (also $\theta_u^{\prime}$ and
$\theta_\gamma^{\prime}$) is not $\pi/2$. The photon direction 
consequently deviate from $(v/c, 0, [1-(v/c)^2]^{1/2})$ slightly. From Fig.2, 
we see that the attenuation coefficient is also very small around the 
direction $(v/c, 0, [1-(v/c)^2]^{1/2})$, so that the conclusion that the 
induced electric field plays a minor role in $\gamma$-ray absorption still 
holds for the oblique rotator case. Actually, DL75's result can only be 
applied to the aligned case strictly, since generally the co-rotating 
velocity ${\bf v}_{r}$ is not equal to 
${\bf v}_{drift}=c({\bf E\times B})/B^2$, with which one can define a frame 
where the electric field vanishes completely. In the co-rotating frame of an 
oblique rotator, an electric field component parallel to the magnetic field 
will still remain so that DL75's application condition fails.

Our results in this paper may have some implications for some previous 
studies which regard the electric field as the important effect of 
$\gamma$-ray absorption (e.g. Hardee 1977; Zhao {\it et al.} 1989; Lu \& 
Shi 1990; Lu, Wei, \& Song 1994; Qiao \& Zhang 1996).

Although the polar cap models of the $\gamma$-ray pulsars are by all means 
sound in principle, their concrete details will alter much by ejecting the 
induced electric field. Specifically, based on DL75's result, Hardee (1977) 
got an absolute upper limit to the photon energies 
$$E_\gamma <9.6\times 10^9 B_{12}^{-1}r_6^2 P {\rm eV}, \eqno(6)$$
(his Eq.(38)) at which escape of the $\gamma$-rays from the magnetosphere 
is possible. It should be replaced by a threshold in the pure magnetic 
field absorption scheme 
$$E_\gamma <2.6\times 10^9 B_{12}^{-1} P^{1/2} {\rm eV} \eqno(7)$$
(Wei, Song, \& Lu 1997, their Eq.(10), $r_6=1$ and the last open field line
is assumed). Thus the generation order parameters proposed by Lu, Wei, \& 
Song (1994) should take the form in Wei, Song, \& Lu (1997, their Eq.(17)).

The three boundary lines (birth line, death line and appearance line) in the 
$\dot P-P$ diagram of pulsars derived by Qiao \& Zhang (1996) are also based 
on the electric field absorption. The details will also be changed by 
ejecting the electric field, but the picture still remains and may give a 
hint to us about the magnetic field configuration in the neutron star 
vicinity (Qiao \& Zhang, discussions).

\acknowledgements{We thank Prof. Lu T., Drs. Song L.M., Wei D.M.,
Zhao Y.H., Xu R.X., and Mr. Liu J.F., Hong B.H. for helpful discussions, 
and Mr. Lin W.P., Dr. Deng L.C. for their help on the figures. BZ 
acknowledges support from the Chinese Post-Doctoral Science Foundation. 
This work is partly supported by NSF of China, the Climbing Project of China 
and the Project Supported by Doctoral Program Foundation of Institution of 
Higher Education in China.}
. 

\end{document}